\documentclass[journal=jpclcd,manuscript=letter]{achemso}

\usepackage{chemformula} 
\usepackage[T1]{fontenc} 
\usepackage{gensymb}
\usepackage[flushleft]{threeparttable}
\usepackage[normalem]{ulem}

\author{Letizia Tavagnacco}
\affiliation{CNR-ISC and Department of Physics, Sapienza University of Rome,\\Piazzale A. Moro 2, 00185, Rome, Italy.}
\author{Ester Chiessi}
\affiliation{Department of Chemical Sciences and Technologies, University of Rome Tor Vergata,\\Via della Ricerca Scientifica I, 00133 Rome, Italy.}
\email{ester.chiessi@uniroma2.it}
\author{Marco Zanatta}
\affiliation{[Department of Computer Science, University of Verona,\\Strada Le Grazie 15, 37138, Verona, Italy.}
\author{Andrea Orecchini}
\affiliation{Department of Physics and Geology, University of Perugia,\\Via A. Pascoli, 06123, Perugia, Italy.}
\alsoaffiliation{CNR-IOM c/o Department of Physics and Geology, University of Perugia,\\Via A. Pascoli, 06123, Perugia, Italy.}
\author{Emanuela Zaccarelli}
\affiliation{CNR-ISC and Department of Physics, Sapienza University of Rome,\\Piazzale A. Moro 2, 00185, Rome, Italy.}
\email{emanuela.zaccarelli@cnr.it}

\title[An \textsf{achemso} demo]
 {Water-polymer coupling induces a dynamical transition in microgels}

\keywords{Microgels, molecular dynamics simulations, dynamical transition, polymer network, hydrogen bonding}

\begin{tocentry}
\centering
\includegraphics[width=5 cm]{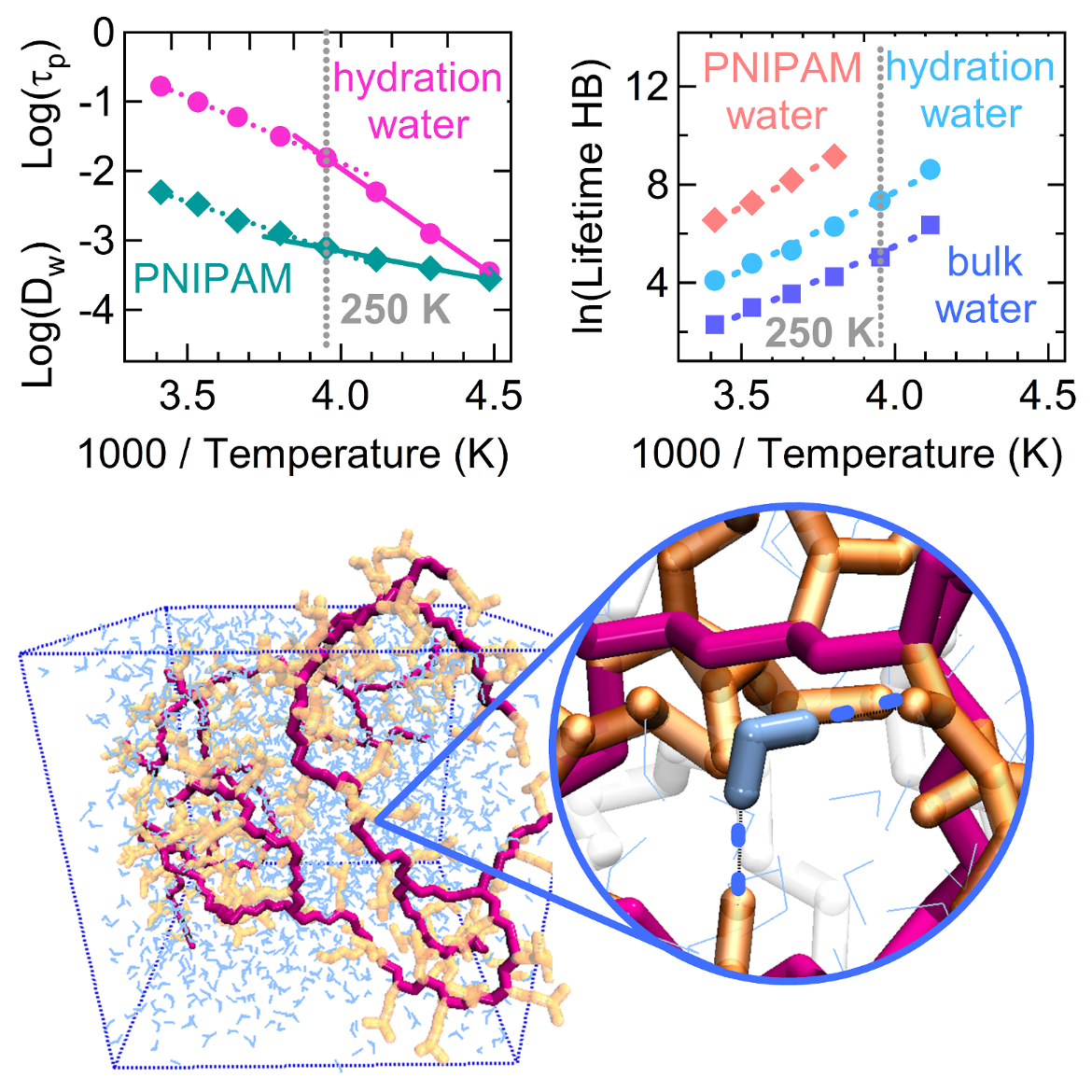}
\end{tocentry}

\begin{document}

\begin{abstract}
The long debated protein dynamical transition was recently found also in non-biological macromolecules, such as poly-N-isopropylacrylamide (PNIPAM) microgels. Here, by using atomistic molecular dynamics simulations, we report a description of the molecular origin of the dynamical transition in these systems.  We show that PNIPAM and water dynamics below the dynamical transition temperature $T_d$ are dominated by methyl group rotations and hydrogen bonding, respectively. By comparing with bulk water, we unambiguously identify PNIPAM-water hydrogen bonding as the main responsible for the occurrence of the transition. The observed phenomenology thus crucially depends on the water-macromolecule coupling, being relevant to a wide class of hydrated systems, independently from the biological function.
\end{abstract}


The well-known protein dynamical transition takes place in hydrated protein suspensions at a relatively low temperature, usually called $T_d$, typically between 220 and 240 K, but depending on the specific system. After the first observation in 1989 for myoglobin~\cite{doster1989dynamical}, the transition has been reported in several kinds of proteins, including ribonuclease A~\cite{rasmussen1992}, cytochrome c~\cite{MARKELZ2007}, lysozyme~\cite{Capaccioli2012} and other biomacromolecules such as RNA~\cite{Caliskan2006}, DNA~\cite{Cornicchi2007} and lipid bilayers~\cite{Peters2017}, irrespective of secondary structure, folding and degree of polymerization~\cite{Roh2005,He2008,schiro2015translational}. The transition consists in a steep enhancement of the atomic mobility  that,  below $T_d$,  is limited to harmonic vibrations and methyl rotations.  At $T_d$ anharmonic motions and local diffusion of groups of atoms are triggered, which, for proteins, is a prerequisite to the onset of activity~\cite{henzler2007dynamic,Dioumaev9621,henzler2007hierarchy}. It is important to remark that $T_d$ does not coincide with the calorimetric glass transition temperature, which is found to occur at even lower temperatures. While the dynamical transition always occurs in aqueous environments, with a water content that is kept to a minimum to avoid the onset of ice crystallization, the role of water is still a debated issue. Evidence was reported in favour of a concomitant activation of the water dynamics at the transition, as shown for folded and intrinsically disordered proteins~\cite{roh2006influence,Chen9012,schiro2015translational}. These studies suggest a strong interplay between protein and water, but different interpretations have also been proposed in the literature going from a water-slaved to a water-driven protein dynamics~\cite{Frauenfelder5129,KHODADADI20101321,Nandi2017}.

Very recently, elastic incoherent neutron scattering experiments reported the occurrence of a dynamical transition at $T_d\sim 250$ K also for a non-biological system, i.e poly(N-isopropyl acrylamide), PNIPAM, microgels~\cite{Zanatta2018}.
Like proteins, PNIPAM microgels have an extended covalent connectivity and an amphipilic character. In addition, their network structure
maximizes the ability of water confinement. For the same macromolecular concentration, water accessible surface area in PNIPAM microgels is about 30\% larger as compared to a globular protein, thus magnifying the matrix-induced effect on water properties\bibnote{The estimate of the water accessible surface area in a PNIPAM microgel at a concentration of 60\% (w/w) and in a hydrated lysozyme sample at the concentration of 67\% (w/w) gives the values of $2.6 \cdot 10^6$ and $2.0 \cdot 10^6$ $m^2 kg^{-1}$, respectively}. This results in the observation of the dynamical transition in PNIPAM-water mixtures with a very large water content, up to roughly 60\% of water (w/w).
These findings thus extend the dynamical transition concept beyond the world of biological function, encompassing the much wider context of hydrated complex macromolecular entities.

To unveil the molecular mechanisms involved in the dynamical transition, in this work we perform atomistic molecular dynamics simulations using a nanoscale model of the microgel in water (Figure \ref{fgr:model}A) that quantitatively reproduces the experimental results~\cite{Zanatta2018}. By varying PNIPAM concentration and exploring a wide range of temperatures, we probe PNIPAM and water dynamics in detail. We find that below $T_d$ both PNIPAM and water motions are characterized by an Arrhenius behavior, controlled by methyl rotations and hydrogen bonds, respectively. By monitoring the hydrogen bonding pattern, we find that long-lived PNIPAM-water bonds are primarily responsible for the observed water behavior at low temperatures. This is confirmed by the comparison with bulk water, which does not display the same features\cite{DeMarzio2016}. Our results clearly highlight the fact that the dynamical transition is a feature genuinely associated to water-macromolecule coupling.

In biological macromolecules, the protein dynamical transition is usually detected by monitoring the  atomic mean squared displacements (MSD) through several experimental techniques~\cite{doster1989dynamical,PARAK1982177,melchers1996structural}. The MSD can also be easily calculated as a function of time by molecular dynamics simulations. Its value at a given time, equal to that of a given experimental resolution, can then be monitored as a function of $T$. This is done in Figure~\ref{fgr:model}B where the MSD of PNIPAM hydrogen atoms of a portion of microgel is reported for both long (1 ns) and short (150 ps) times. On increasing temperature, a sudden promotion of large amplitude motions is visible in both cases, thus excluding the possibility that the observation of a transition is an artefact induced by the resolution limit of the accessible time or frequency window~\cite{Khodadadi2008}.
In addition, the concentration dependence of the MSD reveals that the transition is more pronounced in more hydrated systems, in agreement with what has been found for proteins~\cite{roh2006influence}. This overall qualitative description of the molecular process does not allow to fully understand the microscopic interplay between water and the macromolecule originating this effect. Therefore, in the following, we analyze in detail the dynamical behavior of both PNIPAM and water.

\begin{figure}[H]
\centering
\includegraphics[width=15 cm]{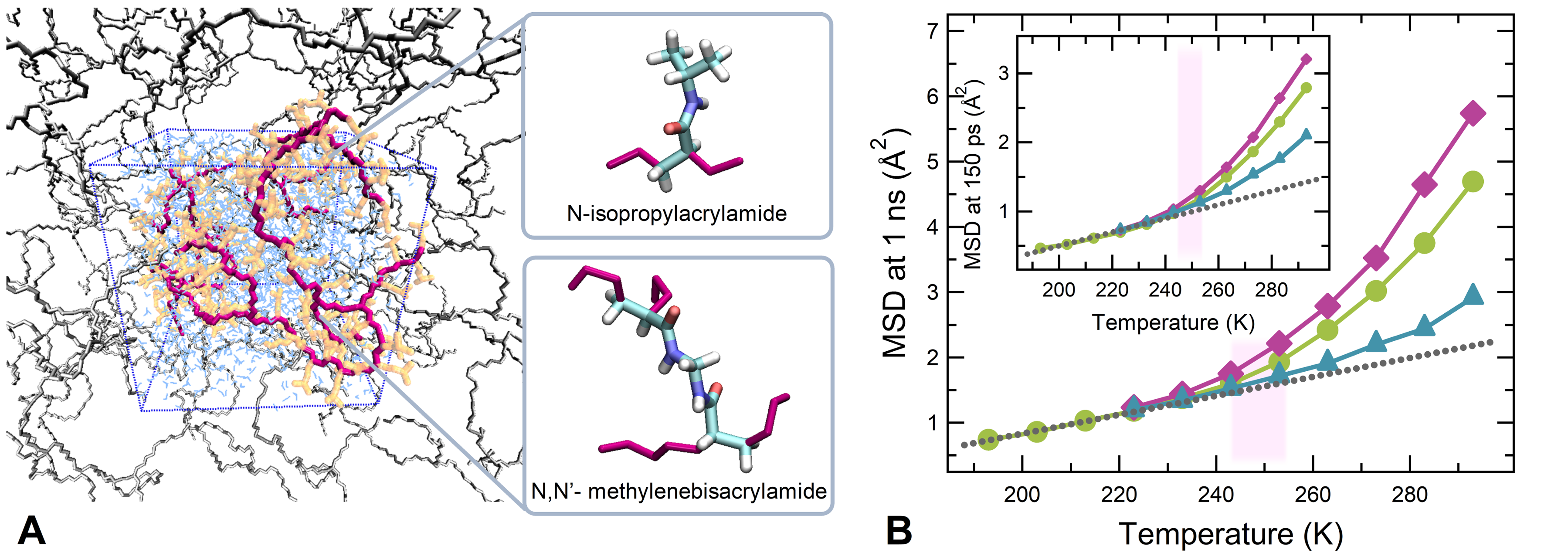}
  \caption{A) Microgel network model. Backbone and side chain atoms are displayed in magenta and yellow, respectively; water molecules and periodic images of backbone atoms are represented in blue and gray, respectively. The two side-frames report the chemical structure of the repeating unit (top) and of the cross-link (bottom). Polymer hydrogen atoms are omitted for clarity; B) Temperature dependence of MSD calculated for PNIPAM hydrogen atoms at 1 ns (main figure) and 150 ps (inset). Results are displayed for PNIPAM mass fractions of 30\% (purple diamonds), 40\% (green circles), and 60\% (blue triangles). The dashed lines are guides to the eye suggesting a linear behavior corresponding to the dry sample, for which the dynamical transition is suppressed~\cite{Zanatta2018}.}
  \label{fgr:model}
\end{figure}

\begin{figure}[H]
\centering
\includegraphics[width=15cm]{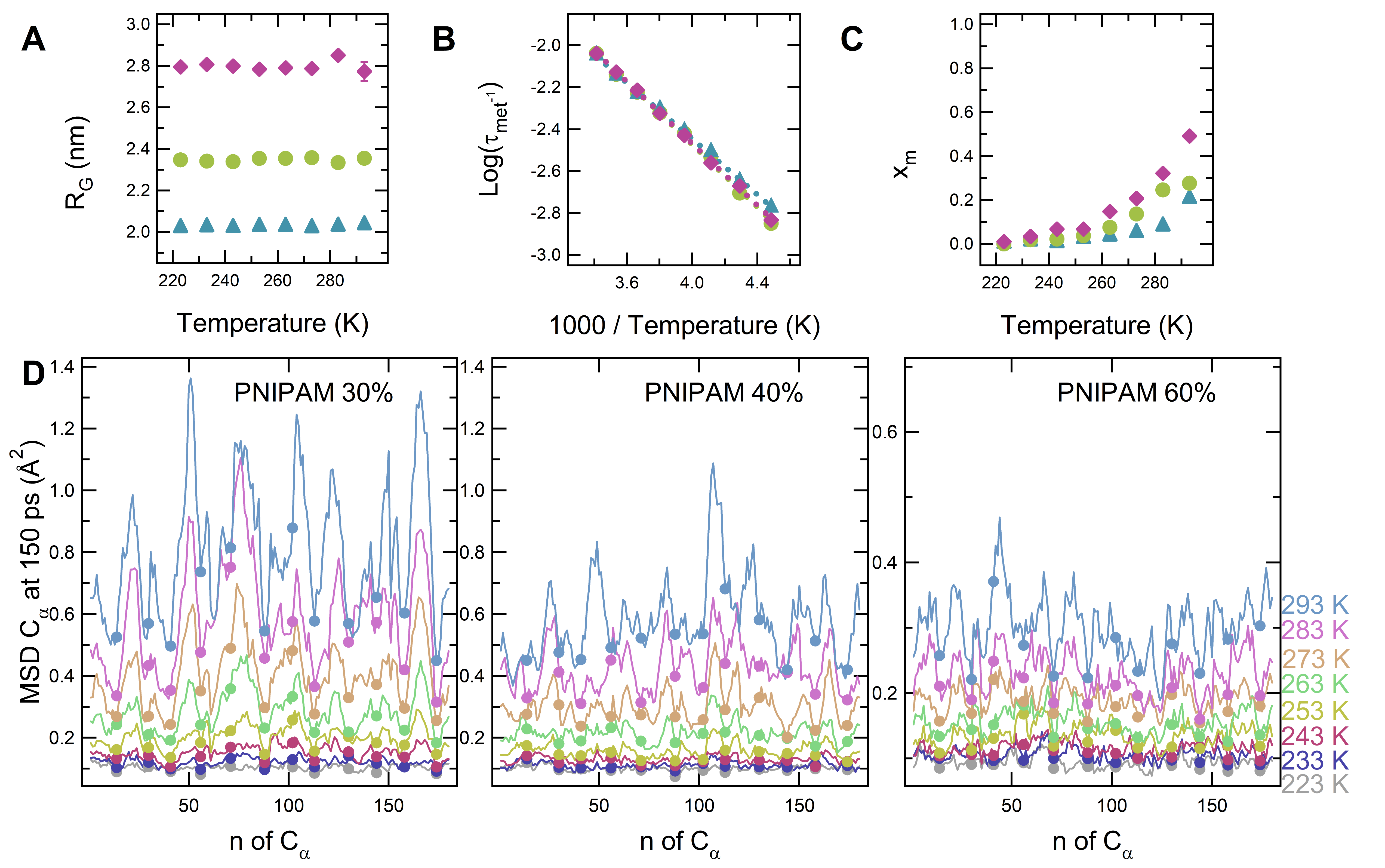}
  \caption{A) Temperature dependence of PNIPAM radius of gyration; B) Arrhenius plot  of the average lifetime (in ps) of a rotational state, $\tau_{met}$, of a methyl group; C) Fraction of backbone mobile dihedrals as a function of the temperature. In A), B) and C) systems with PNIPAM mass fractions of 30, 40 and 60 \% (w/w) are displayed with purple diamonds, green circles and blue triangles, respectively;  D) Temperature dependence of the MSD of backbone tertiary carbon atoms $C_{\alpha}$, that are directly connected to the side chain. Here, $n$ is the index of residue, varying from 1 to 180, the total number of repeating units in our network model. Circles indicate the 12 carbon atoms belonging to crosslinks. The MSD is calculated at 150 ps for PNIPAM mass fraction of 30\% (left panel), 40\% (central panel) and 60\% (right panel). Errors within the symbol size.}
  \label{fgr:structure}
\end{figure}

First of all, it is important to exclude that the transition is due to an underlying structural change in the system.
To this aim, we analyzed the structural alterations of the polymer matrix in the 200$-$290 K interval. Similarly to what is observed in proteins~\cite{doster1989dynamical}, no structural variations of the polymer network were found at $T_d$, neither as a whole, as shown by the temperature dependence of PNIPAM radius of gyration reported in Figure~\ref{fgr:structure}A, nor locally, as visible from the distributions of dihedral angles of the backbone and the temperature dependence of hydrophilic and hydrophobic interactions reported in the Figures S1 and S2, thus confirming the kinetic character of the transition.

The homogeneity of chemical composition in PNIPAM microgels  allows us to identify a hierarchy of motion modes. The expected fastest motion is the torsion of side chain methyl groups, which is active at all the explored temperatures (Figure S3 and Table S1). Indeed by the extending the explored temperature range down to 63 K, an onset of anharmonicity consistent with the activation of methyl group rotation is detected at about 150 K (see Figure S4), in agreement with the behavior observed in biomacromolecules~\cite{Roh2005,schiro2010direct,schiro2010molecular,Telling2011}. We find that the average lifetime of a methyl rotational state is independent on the degree of hydration and follows an Arrhenius behavior (Figure~\ref{fgr:structure}B) for all investigated temperatures with an activation energy, $E_a^{met}$, of about 13.5~kJ~mol$^{-1}$ (Table S3). This value is similar to that reported for methyl groups in the hydrophobic core of proteins, in homomeric polypeptides and in bulk polymers~\cite{Xue2007,Ahumada2002,schiro2010direct}.

Another class of motions in the polymer network can be ascribed to the rotation of the backbone dihedral angles. Figure~\ref{fgr:structure}C shows the evolution of the fraction of mobile backbone dihedral angles $x_m$ with $T$. Its behavior closely resembles that of the MSD behavior of PNIPAM (Fig.~\ref{fgr:model}): the mobile backbone dihedral angles are very scarce below $T_d$ and an abrupt increase of their number occurs at and above $T_d$. The details of the backbone torsional dynamics are reported in the SI text (see Table S2).
A further analysis of atomic motions has been carried out by calculating the MSD of the tertiary carbon atoms of the backbone.  Figure~\ref{fgr:structure}D shows that the segmental dynamics of the polymer scaffold is quenched up to $\sim 250$ K, but is switched on at  higher temperatures.  In addition, while the dynamical behavior of PNIPAM is overall homogenous below $T_d$, above this temperature we observe the onset of dynamical heterogeneities. Namely, the carbon atoms belonging to the junctions of the network (solid points in Figure~\ref{fgr:structure}D) develop lower local diffusivities, evidencing the role of the topological constraints on the dynamics.
The comparison between different PNIPAM concentrations highlights the plasticizing effect of water, which results in a larger mobility for a higher degree of hydration.
These results focus the principal role played by the backbone dynamics in the dynamical transition, in agreement with what observed in experimental studies on polypeptides \cite{schiro2010direct,schiro2010molecular}. Moreover, the enhancement of anharmonic fluctuations at increasing water content (Figures ~\ref{fgr:model}B and ~\ref{fgr:structure}D) is similar to what observed for polypeptides in the presence of side-chains \cite{schiro2010molecular}.

\begin{figure}[H]
\centering
\includegraphics[width=15cm]{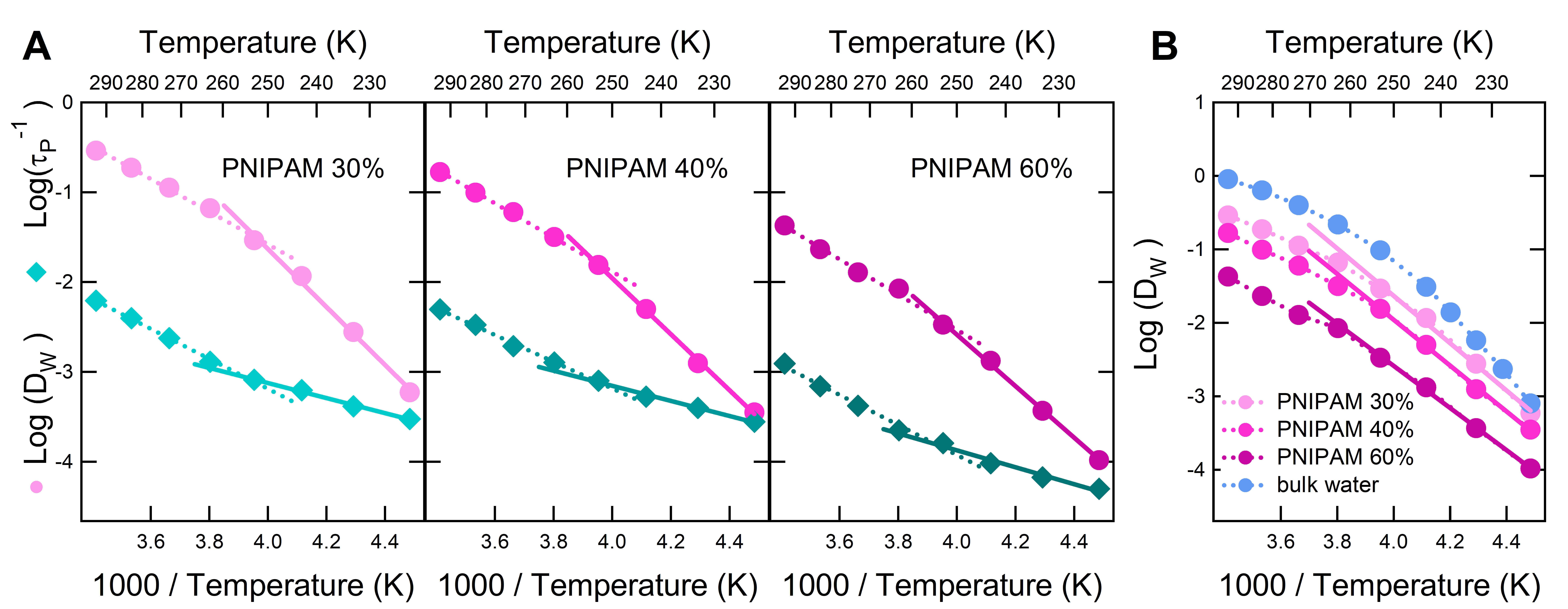}
  \caption{Arrhenius plots of A) water diffusion coefficient $D_w$ (circles, in cm$^2/$s $\times 10^5$) and SISF relaxation times for PNIPAM hydrogen atoms $\tau_p$ (diamonds, in ps) calculated for PNIPAM mass fraction of 30\% (left panel), 40\% (central panel) and 60\% (right panel); B) water diffusion coefficient (in cm$^2/$s $\times 10^5$) calculated for bulk water and hydration water for different PNIPAM concentrations. In all panels, solid lines are Arrhenius fits to the data and dotted lines are guides to the eye.}
  \label{fgr:Ea}
\end{figure}

\begin{figure}[H]
\centering
\includegraphics[width=12cm]{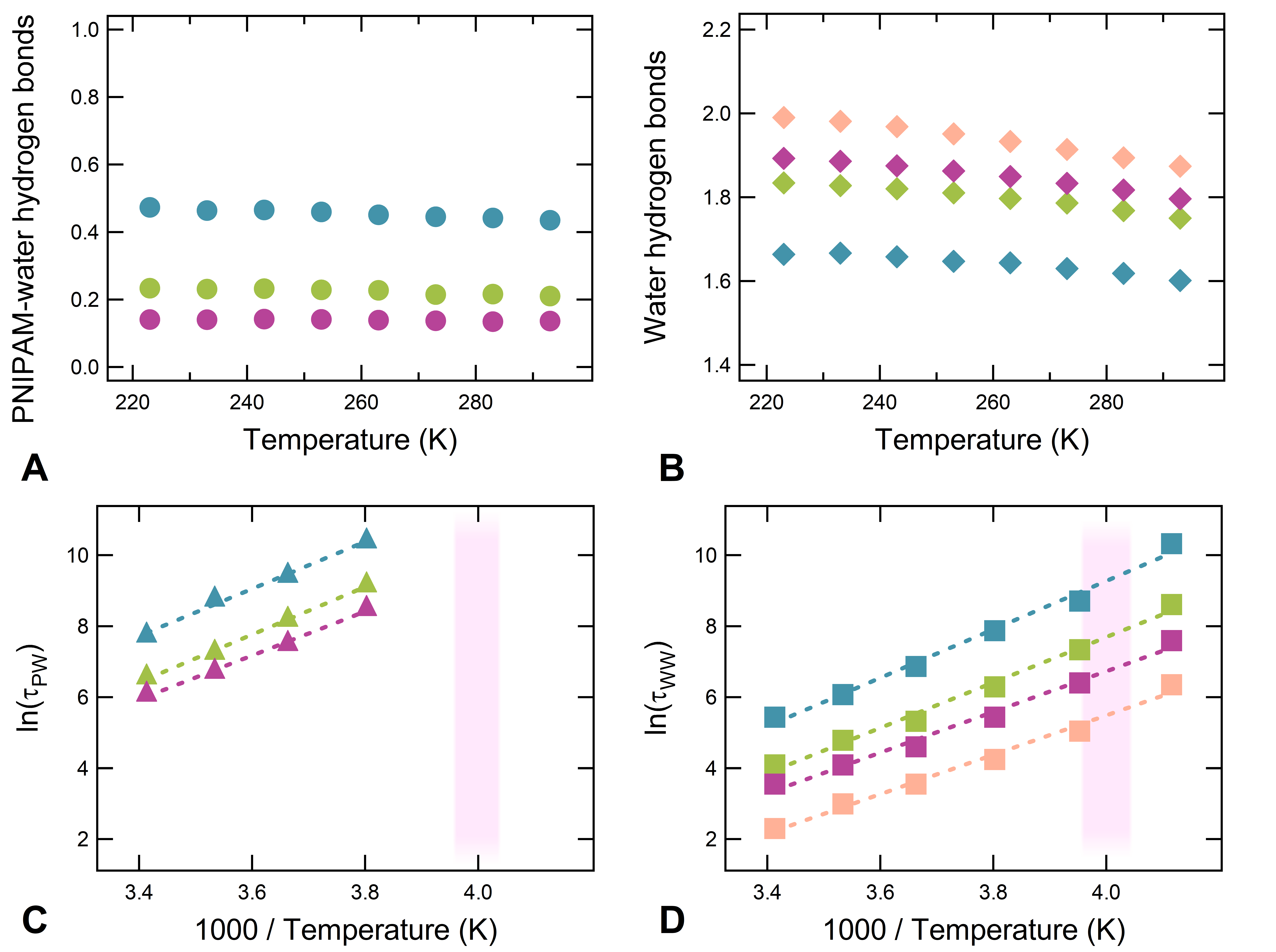}
  \caption{Temperature dependence of the average number of hydrogen bonds between PNIPAM and water (A) and
  between water molecules (B) \emph{per} water molecule; Arrhenius plot of the average lifetime (in ps) of PNIPAM-water hydrogen bonds $\tau_{PW}$ (C) and  of water-water hydrogen bond $\tau_{WW}$ (D). Results referring to PNIPAM mass fraction of 30\%, 40\% and 60\% wt are displayed in purple, green, and blue, respectively. Bulk water data are shown in pink. The pink frames highlight the region where the dynamical transition occurs. Errors within the symbol size.}
  \label{fgr:HB}
\end{figure}

In order to monitor the PNIPAM dynamics as a function of $T$, we consider the self intermediate scattering function (SISF) of PNIPAM hydrogen atoms (Table S4 and SI text), which is well described, at long times, by a stretched exponential with characteristic time $\tau_p$.
The SISF is sensitive to the single particle dynamics at a specific wavevector $Q$. We focus on the value $Q=2.25$\AA$^{-1}$, corresponding, in real space, to the position of the first maximum of the oxygen-oxygen structure factor in bulk water~\cite{SciortinoPRE}.
In Figure~\ref{fgr:Ea}A an Arrhenius plot of $\tau_{p}$ is reported for three different PNIPAM concentrations, indicating a slowing down at the highest PNIPAM concentration, while the systems at 30\% and 40\% (w/w) of polymer display similar values of $\tau_p$ (Table S3). We also report in Figure~\ref{fgr:Ea}A the water translational self-diffusion coefficient $D_w$ as a function of $T$. The direct comparison between PNIPAM and water dynamics allows us to highlight several important features. First of all, we notice that both $\tau_p$ and $D_w$ display a  change of behavior around $\sim 250$ K, that we identify as the dynamical transition.
Analysing the behaviors of $\tau_p$ and $D_w$ in detail, we notice that they are compatible with an Arrhenius dependence both below and above $T_d$\cite{Zanatta2018}. However, we will discuss later that only the low-temperature regime can be attributed to a true activation process. Instead, the high temperature data show a remarkably similar $T$-dependence for both $\tau_{p}$ and $D_w$ at all explored concentrations with a common apparent Arrhenius slope. We further notice that an inversion of the (putative) Arrhenius slope is detected between PNIPAM and water across $T_d$. In particular, for PNIPAM there is a crossover between a lower and a higher slope regime when temperature is increased above $T_d$, while water shows the opposite trend.
\\
We recall that the number of hydration water molecules for PNIPAM is experimentally estimated to be $\sim12\pm1$ per residue below the lower critical solution temperature~\cite{Ono2006,Satokawa2009}. Therefore, for the three concentration values considered in this work, the water molecules can be entirely classified as hydration water (see also Methods).  Hence, it is instructive to compare the $T$-dependence of $D_w$ between hydration water and bulk water, as shown in Figure~\ref{fgr:Ea}B.  No clearly identifiable change at $T_d$ occurs for bulk water, suggesting the central role of PNIPAM-water coupling in the occurrence of the dynamical transition.
Having identified the presence of a dynamical coupling between PNIPAM and water, we now ask the question whether a change of connectivity for the water molecules occurs at the transition. To this aim, we report in Figure~\ref{fgr:HB}A,B the temperature behavior of the number of PNIPAM-water and water-water hydrogen bonds (HB), respectively. None of these observables display a sharp variation at $T_d$.  We find an increasing number of PNIPAM-water HB as a function of PNIPAM concentration and concomitantly a decrease of the number of water-water HB, with no discontinuity with respect to the behavior of bulk water. This suggests that no increase of water structuring is originated in the surrounding of hydrophobic groups.
It is important to stress that the overall water structuring in microgel suspensions is always larger than for bulk water, with the total number of HB (water-water and water-PNIPAM) increasing with polymer concentration (Figure S5). Thus, we can discard a structural origin of the transition in terms of HB connectivity.

\noindent We further monitor the characteristic lifetime of PNIPAM-water $\tau_{PW}$  and water-water $\tau_{WW}$ HB interactions. As shown in Figures \ref{fgr:HB}C and \ref{fgr:HB}D, $\tau_{PW}$ and $\tau_{WW}$ both follow an Arrhenius behavior with similar values of activation energies of about 55~kJ~mol$^{-1}$ (Table S3), irrespective of  hydration level. We also note that the lifetime of PNIPAM-water HB is considerably longer than the lifetime of water-water HB and cannot be estimated below $T_d$ due to the finite simulation time. This finding is in agreement with previous results on the protein dynamical transition~\cite{Tarek2002}, which suggested that the lifetime of protein-water HB interactions diverges at $T_d$.
The activation energy of HB is equal to that estimated from the Arrhenius dependence of $D_w$ for $T<T_d$ (see Table S3).
Instead, the PNIPAM relaxation time $\tau_p$ has an activation energy of $\sim16$~kJ~mol$^{-1}$ for $T<T_d$ which is independent on the concentration and is comparable to that calculated for the rotation of the methyl groups (see Figure~\ref{fgr:structure}B).
These findings suggest that below $T_d$ PNIPAM dynamics is dominated by methyl rotations, while the translational dynamics of water is controlled by its hydrogen bonding within the microgel environment.

\noindent For $T > T_d$,  both observables $\tau_p$ and $D_w$ show a similar temperature dependence, that could be interpreted as an Arrhenius regime\cite{Zanatta2018}, providing activation energies very close to each other, i.e. $\sim30\pm2$~kJ~mol$^{-1}$ for  $\tau_p$  and $\sim36\pm2$~kJ~mol$^{-1}$ for $D_w$. However, these values cannot be attributed to any specific structural rearrangement, indicating that this apparent Arrhenius behavior\cite{Kanaya1999} may result from the superposition of different contributions. The fact that the apparent activation energy is higher than the methyl rotation one and lower than the one corresponding to HB lifetime gives rise to a ``slope'' inversion of PNIPAM and water dynamics, as observed in the data of Fig.~\ref{fgr:Ea}, similarly to hydrated lysozyme results\cite{camisasca2016two}. Hence, above the dynamical transition temperature, we cannot distinguish a dominant specific molecular motion, but rather we only detect a correlation between polymer segmental dynamics and diffusive motion of bound water. This dynamical interplay can only be active when the PNIPAM-water HB lifetime is sufficiently short that the exchange of hydrogen bonded water molecules is still effective~\cite{Tarek2002} and was recently confirmed in the water soluble states of a PNIPAM linear chain~\cite{Tavagnacco2018}.

\noindent Differently, the dynamical transition clearly marks the onset, below $T_d$, of distinct dominant motions, respectively for PNIPAM and water.  The comparison with bulk water further allows us to identify that it is primarily the hydrogen bond pattern of water within the microgel, dictated by the slowest PNIPAM-water HB, that dominates the low-T water dynamics. The $T$-dependence of $D_w$ for bulk water is further analyzed in the Supplementary Information, where it is shown that at high temperatures the data are well described by a power-law decay (see Figure S6 Supplementary Information) in agreement with previous studies\cite{camisasca2016two,DeMarzio2016}. For low $T$, also bulk water follows an Arrhenius dependence but with a much higher activation energy ($\sim80$~kJ~mol$^{-1}$), indicating a different mechanism than hydrogen bonding controlling its dynamics, even if the number of water-water HB increases as $T$ decreases. This leads us to exclude a direct influence of water-water hydrogen bonding on the dynamical transition. Instead, we can clearly isolate the contribution of PNIPAM-water hydrogen bonding, which is slower than water-water HB by roughly two orders of magnitude at each $T$ for the same PNIPAM concentration (see Figures~\ref{fgr:HB}C and D). Therefore, PNIPAM-water hydrogen bonding must be the trigger of the Arrhenius dependence below $T_d$ and thus the microscopic  mechanism responsible for the occurrence of the dynamical transition.

\noindent  It is also interesting to discuss the relatively large value of $T_d$ for PNIPAM microgels with respect to protein systems.
The average number of macromolecule-water hydrogen bonds formed was found to be 1.2 and 1.1 for lysozyme and plastocyanin \emph{per} protein residue, respectively\cite{Nandi2017,arcangeli1998role}, while for PNIPAM-water hydrogen bonds we find $\sim 2$\bibnote{For PNIPAM 40\% mass fraction, the number of PNIPAM-water hydrogen bonds normalized to the number of PNIPAM residues is obtained by multiplying the value reported in the Figure~\ref{fgr:HB}A to the number of water molecules (1676) and dividing it by the number of PNIPAM residues (180).}. Thus, the different macromolecule-water interaction may play a primary role in determining the value of $T_d$. We also note that the value of $T_d$ was also found to depend on the hydration level $h$\cite{paciaroni2005fast,Capaccioli2012,kim2016computational} ($h$ = g water / g protein), with lower values found for higher water content. However, such a variation appears to be significant at low $h$ \cite{paciaroni2005fast}, tending to saturate at large water content.  We stress that the PNIPAM mass fractions that we have explored correspond to water content $h=2.33, 1.5, 0.67$  for 30\%, 40\% and 60\%, respectively, thus to a completely different water content region with respect to all previous studies.

In conclusion, the present results provide a microscopic description of the origin of the ``protein-like'' dynamical transition observed in microgels.
The correlation between the information extracted from the analysis of PNIPAM relaxations times, water self-diffusion coefficients and hydrogen bonding interactions allowed us to identify the molecular processes which control the dynamics of both PNIPAM and water below the dynamical transition temperature. In particular, we found that below $T_d$ PNIPAM dynamics is governed by the rotation of the methyl groups belonging to the side chains and that a sudden increase of the polymer segmental dynamics occurs at $T_d$. On the other hand, hydrogen bonding interactions determine water dynamics below $T_d$. By comparing the low temperature behavior of water in the microgels suspensions to that of bulk water, we also found that the hydrogen bonding interactions between PNIPAM and water play the primary role in determining water dynamics below $T_d$.  Altogether these findings support the idea that the macromolecular-water coupling is the driving ingredient of the dynamical transition. Thus, such phenomenology should be rather general, taking place in all hydrated macromolecular systems which are able, at the same time, to efficiently confine water in order to avoid ice formation and to couple it via hydrogen bonding. These findings also rule out the possibility to observe a dynamical transition in dry systems, where similar findings \cite{MamontovPRL} must then be ascribed to a different molecular mechanism.

\section{Experimental}
PNIPAM microgel is modelled as an isotropic network composed by 12 atactic chains connected by 6 bisacrylamide cross-links. Periodic images of the network are covalently bonded to mimic the 3-D percolation of the microgel. This model has a monomer/cross-linker ratio that, given the heterogeneous structure of PNIPAM microgels, describes a region of the particle close to the core-shell boundary and it quantitatively reproduces neutron scattering results~\cite{Zanatta2018}. Three PNIPAM mass fractions of 30, 40 and 60\% (w/w), corresponding to hydration levels of 14, 9 and 4 water molecules per PNIPAM residue, respectively, were investigated. All-atom MD simulations are performed in the range between 293~K and 223~K. At each $T$, trajectory data are collected for $\sim 0.5~\mu$s. We adopt the OPLS-AA force field~\cite{Jorgensen1996} with the implementation by Siu et al.~\cite{Siu2012} for PNIPAM and the Tip4p/ICE model~\cite{tip4pICE} for water. MD simulations are also carried out for a cubic box containing 1782 Tip4p/ICE water molecules following a similar procedure. Additional details are reported in sections S7-S9 of the SI Text.

\begin{acknowledgement}
LT, EC and EZ acknowledge support from the European Research Council (ERC Consolidator Grant 681597, MIMIC) and CINECA-ISCRA for HPC resources.
\end{acknowledgement}

\begin{suppinfo}
The Supporting Information is available free of charge with the following sections:

\begin{itemize}
  \item S1. Study of PNIPAM microgel structural rearrangements
  \item S2. Torsional dynamics
  \item S3. Activation energies
  \item S4. Self intermediate scattering functions: fitting procedure
  \item S5. Water hydrogen bonding
  \item S6. Bulk water diffusion coefficient
  \item S7. \emph{In silico} model of PNIPAM microgel
  \item S8. Simulation protocol
  \item S9. Data analysis
\end{itemize}
\end{suppinfo}


\begin{mcitethebibliography}{53}
\bibitem[Not()]{Note-2}
For PNIPAM 40\% mass fraction, the number of PNIPAM-water hydrogen bonds
  normalized to the number of PNIPAM residues is obtained by multiplying the
  value reported in the Figure~\ref{fgr:HB}A to the number of water molecules
  (1676) and dividing it by the number of PNIPAM residues (180).\relax
\mciteBstWouldAddEndPunctfalse
\mciteSetBstMidEndSepPunct{\mcitedefaultmidpunct}
{}{\mcitedefaultseppunct}\relax
\EndOfBibitem
\end{mcitethebibliography}

\newpage
\renewcommand{\thefigure}{S\arabic{figure}}
\renewcommand{\thetable}{S\arabic{table}}
\setcounter{figure}{0}

\section{Supporting Information}

\section{S1. Study of PNIPAM microgel structural rearrangements}
\begin{figure}[H]
\centering
\includegraphics[width=12 cm]{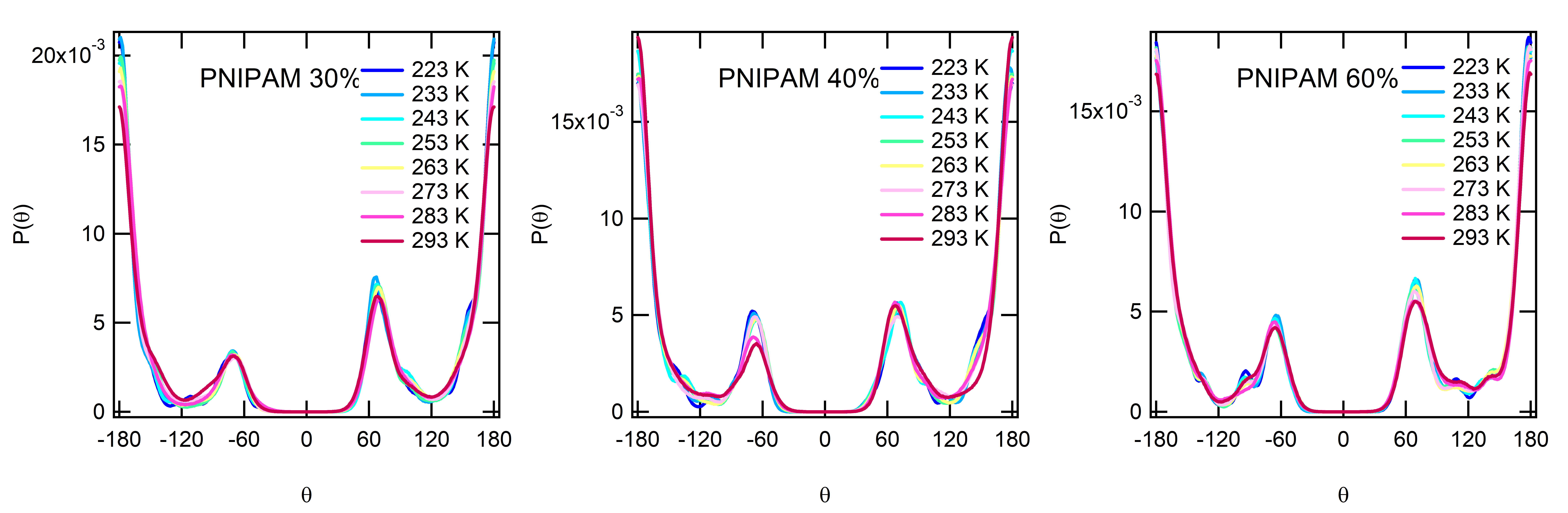}
  \caption{Distribution of backbone dihedral angles of PNIPAM as a function of the temperature calculated over 200 ns. Results referring to the system with a PNIPAM mass fraction of 30\% (w/w) are displayed on the left, 40\% (w/w) in the middle, and 60\% (w/w) on the right. }
  \label{fgr:BBdist}
\end{figure}

\begin{figure}[H]
\centering
\includegraphics[width=6 cm]{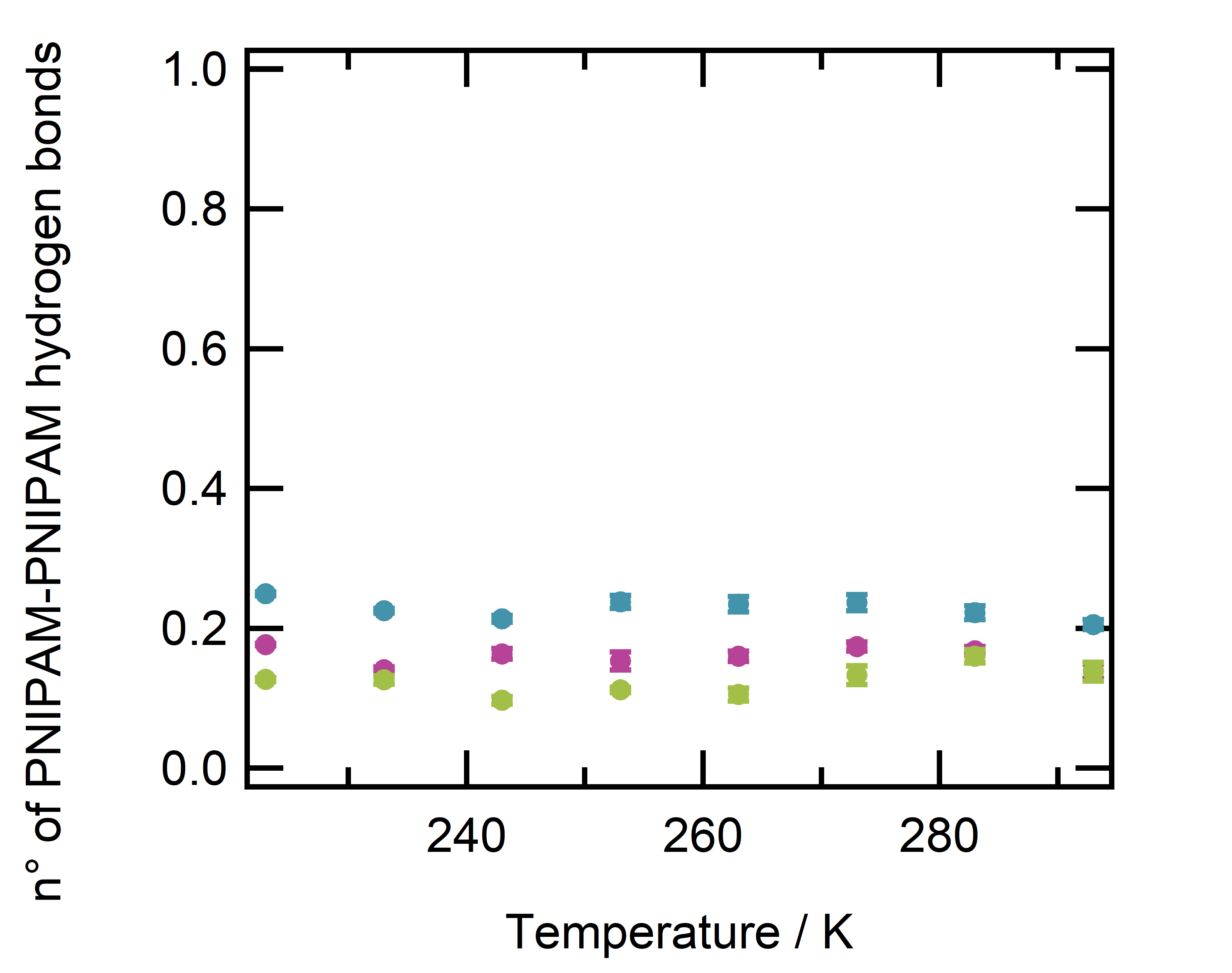}
  \caption{Temperature dependence of the average number of hydrogen bonds between PNIPAM amide groups normalized to the number of repeating units of the polymeric network, calculated over the last 100 ns. Results referring to the system with PNIPAM mass fraction of 30\%, 40\% and 60\% (w/w) are displayed in purple, green, and blue, respectively. Errors within symbol size.}
  \label{fgr:PPhb}
\end{figure}

We investigated the presence of local structural rearrangements in the polymer network by monitoring the temperature dependence of the distributions of backbone dihedral angles. At all the explored concentrations, the backbone dihedral distributions do not show a clear variation at $T_d$ (Figure \ref{fgr:BBdist}). In addition, we probed how intramolecular hydrophilic and hydrophobic interactions changed with the temperature. The average number of hydrogen bonds between PNIPAM amide groups has no clear variation at $T_d$, for all the investigated systems, as shown in Figure \ref{fgr:PPhb}. Similarly, the average number of hydrophobic contacts is constant with the temperature to a value of 2 \emph{per} repeating unit of the polymer network.

\section{S2. Torsional dynamics}
\begin{figure}[H]
\centering
\includegraphics[width=11 cm]{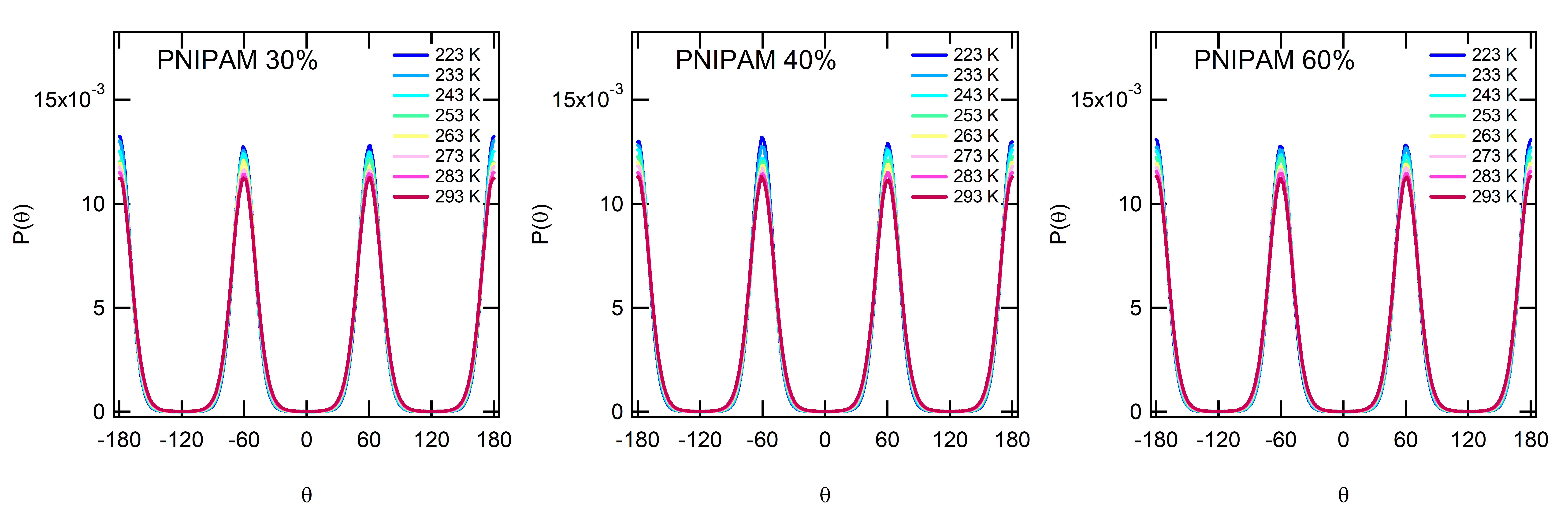}
  \caption{Distribution of dihedral angles of the methyl groups in the side chain of PNIPAM as a function of the temperature calculated over the last 100 ns. Results referring to the system with a PNIPAM mass fraction of 30\% (w/w) are displayed on the left, 40\% (w/w) in the middle, and 60\% (w/w) on the right. Errors within symbol size.}
  \label{fgr:METdist}
\end{figure}

\begin{figure}[H]
\centering
\includegraphics[width=11 cm]{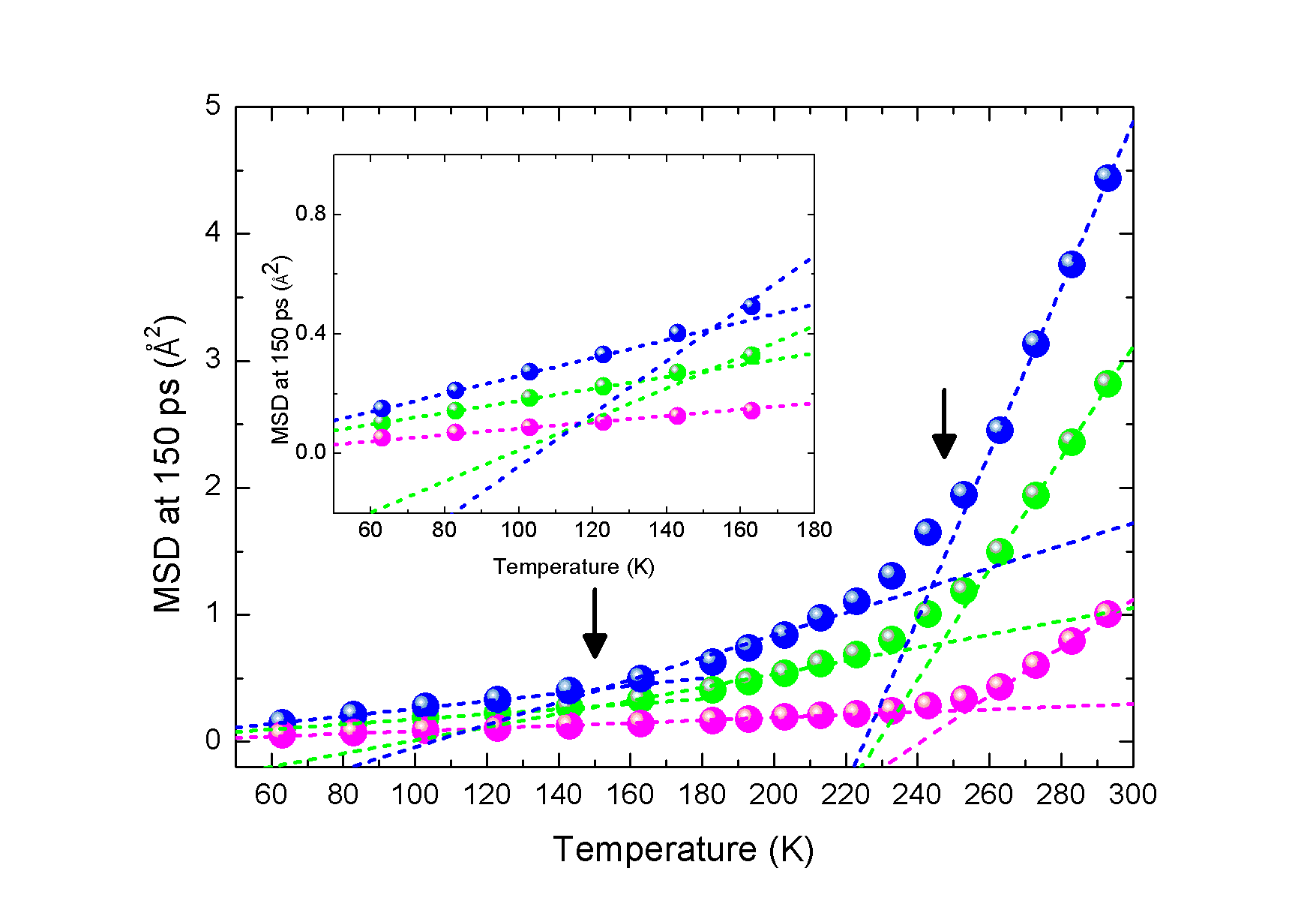}
  \caption{Temperature dependence of the mean squared displacement of PNIPAM hydrogen atoms calculated at 150 ps for the system with a PNIPAM mass fraction of 40\% (w/w). Data calculated for all hydrogen atoms, hydrogen atoms belonging to the methyl groups and hydrogen atoms not belonging to the methyl groups are shown in green, blue and pink, respectively. Dotted lines are guide to the eye. The inset is a zoom of the low temperature region.}
  \label{fgr:MSDmet}
\end{figure}

We studied the onset of anharmonic motions in the polymer network by investigating the conformation and torsional dynamics of the backbone and the methyl groups belonging to PNIPAM side chains. The distributions of dihedral angles of the methyl groups show a uniform behavior with the temperature, with conformational states equally populated (Figure \ref{fgr:METdist}). In addition, as reported in Table \ref{tbl:methyl}, transitions between conformational states of methyl groups are detected at each temperature and all the dihedrals are active in the whole temperature range. On the contrary, in the case of backbone dihedrals, a clear increase of the number of mobile dihedrals and dihedrals transitions takes place at $T_d$ (see Table \ref{tbl:backbone}). For PNIPAM 40\% w/w we also extended the investigated temperature range down to 63 K and we evaluated independently the MSD contribution of PNIPAM hydrogen atoms belonging and not to the side chain methyl groups, as reported in Figure \ref{fgr:MSDmet}. We notice that the dynamics of hydrogen atoms not belonging to methyl groups is harmonic up to about $\sim 250$ K, while the MSD of hydrogen atoms of methyl groups deviates from a linear dependence at $\sim 150$ K. The low temperature anharmonic onset can thus be related to the activation of methyl groups rotations.

\begin{table}[H]
  \begin{threeparttable}
  \caption{Torsional dynamics of the dihedrals of the methyl groups.}
  \label{tbl:methyl}
  \begin{tabular}{cccccccccc}
    \hline
                       &\multicolumn{3}{c}{\textbf{PNIPAM 30\%}}&\multicolumn{3}{c}{\textbf{PNIPAM 40\%}}& \multicolumn{3}{c}{\textbf{PNIPAM 60\%}} \\
    \hline
    T (K) & $n_t$ & $\tau_{met}$ (ps) & $x_m$ & $n_t$ & $\tau_{met}$ (ps) & $x_m$ & $n_t$ & $\tau_{met}$ (ps) & $x_m$ \\
    \hline
    223  & 46099  & 6.8 $\cdot 10^2$ & 100 & 44452  & 7.1 $\cdot 10^2$ & 100 & 52912  & 5.9 $\cdot 10^2$ & 100 \\
    233  & 67052  & 4.7 $\cdot 10^2$ & 100 & 61966  & 5.1 $\cdot 10^2$ & 100 & 70471  & 4.5 $\cdot 10^2$ & 100 \\
    243  & 86459  & 3.6 $\cdot 10^2$ & 100 & 92121  & 3.4 $\cdot 10^2$ & 100 & 97445  & 3.2 $\cdot 10^2$ & 100 \\
    253  & 116837 & 2.7 $\cdot 10^2$ & 100 & 119590 & 2.6 $\cdot 10^2$ & 100 & 122006 & 2.6 $\cdot 10^2$ & 100 \\
    263  & 148676 & 2.1 $\cdot 10^2$ & 100 & 149880 & 2.1 $\cdot 10^2$ & 100 & 155509 & 2.0 $\cdot 10^2$ & 100 \\
    273  & 191875 & 1.6 $\cdot 10^2$ & 100 & 187658 & 1.7 $\cdot 10^2$ & 100 & 184050 & 1.7 $\cdot 10^2$ & 100 \\
    283  & 234109 & 1.3 $\cdot 10^2$ & 100 & 228333 & 1.4 $\cdot 10^2$ & 100 & 225295 & 1.4 $\cdot 10^2$ & 100 \\
    293  & 287349 & 1.1 $\cdot 10^2$ & 100 & 287702 & 1.1 $\cdot 10^2$ & 100 & 281009 & 1.1 $\cdot 10^2$ & 100 \\
    \hline
    \end{tabular}
        \begin{tablenotes}
        \footnotesize
        \item \emph{$T$ is the temperature, $n_t$ is the number of dihedral transitions, $\tau_{met}$ is the average lifetime of a dihedral rotational state (error within the last significant digit) and $x_m$ is the percentage of mobile dihedrals. Analysis over the last 100 ns.}
        \end{tablenotes}
    \end{threeparttable}
\end{table}

\begin{table}[H]
  \begin{threeparttable}
  \caption{Torsional dynamics of PNIPAM backbone dihedrals.}
  \label{tbl:backbone}
  \begin{tabular}{cccccccccc}
    \hline
                       &\multicolumn{3}{c}{\textbf{PNIPAM 30\%}}&\multicolumn{3}{c}{\textbf{PNIPAM 40\%}}& \multicolumn{3}{c}{\textbf{PNIPAM 60\%}} \\
    \hline
    T (K) & $n_t$ & $\tau_{bb}$ (ns) & $x_m$ & $n_t$ & $\tau_{bb}$ (ns) & $x_m$ & $n_t$ & $\tau_{bb}$ (ns) & $x_m$ \\
    \hline
    223  & 8    & $-$ & 1.1 & 0    & $-$ & $-$ & 0   & $-$ & $-$ \\
    233  & 48   & $-$ & 3.4 & 20   & $-$ & 1.9 & 34  & $-$ & 1.1 \\
    243  & 76   & $-$ & 6.8 & 21   & $-$ & 2.3 & 8   & $-$ & 0.4 \\
    253  & 97   & 55  & 6.8 & 74   & 71  & 3.8 & 20  & 260 & 2.3 \\
    263  & 346  & 15  & 15  & 138  & 38  & 7.6 & 23  & 230 & 3.4 \\
    273  & 733  & 7.2 & 21  & 341  & 15  & 14  & 85  & 62  & 4.9 \\
    283  & 1083 & 4.9 & 32  & 617  & 8.6 & 25  & 256 & 21  & 7.9 \\
    293  & 1945 & 2.7 & 49  & 1003 & 5.3 & 28  & 560 & 9.4 & 20  \\
    \hline
    \end{tabular}
        \begin{tablenotes}
        \footnotesize
        \item \emph{$T$ is the temperature, $n_t$ is the number of dihedral transitions, $\tau_{bb}$ is the average lifetime of a dihedral rotational state (error within the last significant digit) and $x_m$ is the percentage of mobile dihedrals. Analysis over the last 200 ns.}
        \end{tablenotes}
    \end{threeparttable}
\end{table}

\section{S3. Activation energies }
\begin{table}[H]
  \begin{threeparttable}
  \caption{Activation energies (kJ~mol$^{-1}$).}
  \label{tbl:EA}
  \begin{tabular}{lccc}
    \hline
                       &\textbf{PNIPAM 30\%}  &  \textbf{PNIPAM 40\%}  &  \textbf{PNIPAM 60\%} \\
    \hline
    \textbf{$\tau_p$}  & $16(\pm1)$            &  $14.5(\pm0.5)$         &  $18(\pm2)$ \\
    \textbf{$D_w$}     & $60(\pm2)$            &  $60(\pm2)$             &  $55(\pm2)$ \\
    \hline
    \textbf{$\tau_{met}$}       &$14.1(\pm0.2)$    &  $14.3(\pm0.3)$  &  $12.8(\pm0.2)$ \\
    \textbf{$\tau_{PW}$}        &$52(\pm2)$       &  $56(\pm2)$      &  $55(\pm3)$ \\
    \textbf{$\tau_{WW}$}        &$52(\pm3)$       &  $56(\pm2)$      &  $57(\pm3)$ \\
    \hline

    \end{tabular}
        \begin{tablenotes}
        \footnotesize
        \item Table summarizing the activation energies associated to different observables: the relaxation time of the SISFs of PNIPAM hydrogen atoms $\tau_p$, the water self-diffusion coefficient $D_w$, the lifetime of the rotational state of a dihedral of a methyl group $\tau_{met}$,  the lifetime of hydrogen bonds between PNIPAM and water molecules $\tau_{PW}$ and between water molecules $\tau_{WW}$.
        \end{tablenotes}
    \end{threeparttable}
\end{table}

\section{S4. Self intermediate scattering functions: fitting procedure }
The dynamical behavior of PNIPAM was investigated by calculating the self intermediate scattering function, SISF, of PNIPAM hydrogen atoms, defined as:
\begin{eqnarray}
F_{self}(Q,t)&=&\frac{1}{N} \langle \sum\limits_{i=1}^N e^{-i \vec{Q} \cdot [ \vec{r}_i(t) - \vec{r}_i(0) ] } \rangle
\label{Eq:SISF}
\end{eqnarray}
The SISFs were calculated at the wave vector Q=2.25 \AA$^{-1}$.
The long time behavior has been fitted with a stretched exponential function:
\begin{eqnarray}
F_{self}(Q,t)&=&f_a \cdot e^{(-t/\tau_p)^\beta}
\label{Eq:stretched}
\end{eqnarray}
from which $\tau_p$ is estimated.
\begin{table}[H]
  \begin{threeparttable}
  \caption{Relaxation time of SISFs calculated on PNIPAM hydrogen atoms.}
  \label{tbl:SISF}
  \begin{tabular}{cccc}
    \hline
                       & \textbf{PNIPAM 30\%} & \textbf{PNIPAM 40\%} & \textbf{PNIPAM 60\%} \\
    \hline
    T (K) & $\tau_{p}$ (ps)  & $\tau_{p}$ (ps) & $\tau_{p}$ (ps) \\
    \hline
    223  & 3.3 $\cdot 10^3 $  & 3.6 $\cdot 10^3 $  & 2.0 $\cdot 10^4 $  \\
    233  & 2.4 $\cdot 10^3 $  & 2.5 $\cdot 10^3 $  & 1.5 $\cdot 10^4 $  \\
    243  & 1.6 $\cdot 10^3 $  & 1.9 $\cdot 10^3 $  & 1.0 $\cdot 10^4 $  \\
    253  & 1.2 $\cdot 10^3 $  & 1.3 $\cdot 10^3 $  & 6.2 $\cdot 10^3 $  \\
    263  & 7.7 $\cdot 10^2 $  & 7.9 $\cdot 10^2 $  & 4.5 $\cdot 10^3 $  \\
    273  & 4.2 $\cdot 10^2 $  & 5.2 $\cdot 10^2 $  & 2.4 $\cdot 10^3 $  \\
    283  & 2.5 $\cdot 10^2 $  & 3.0 $\cdot 10^2 $  & 1.4 $\cdot 10^3 $  \\
    293  & 1.6 $\cdot 10^2 $  & 2.0 $\cdot 10^2 $  & 8.1 $\cdot 10^2 $  \\
    \hline
    \end{tabular}
        \begin{tablenotes}
        \footnotesize
        \item \emph{Analysis over the last 10 ns. Error within the last significant digit.}
        \end{tablenotes}
    \end{threeparttable}
\end{table}

\section{S5. Water hydrogen bonding}
\begin{figure}[H]
\centering
\includegraphics[width=5 cm]{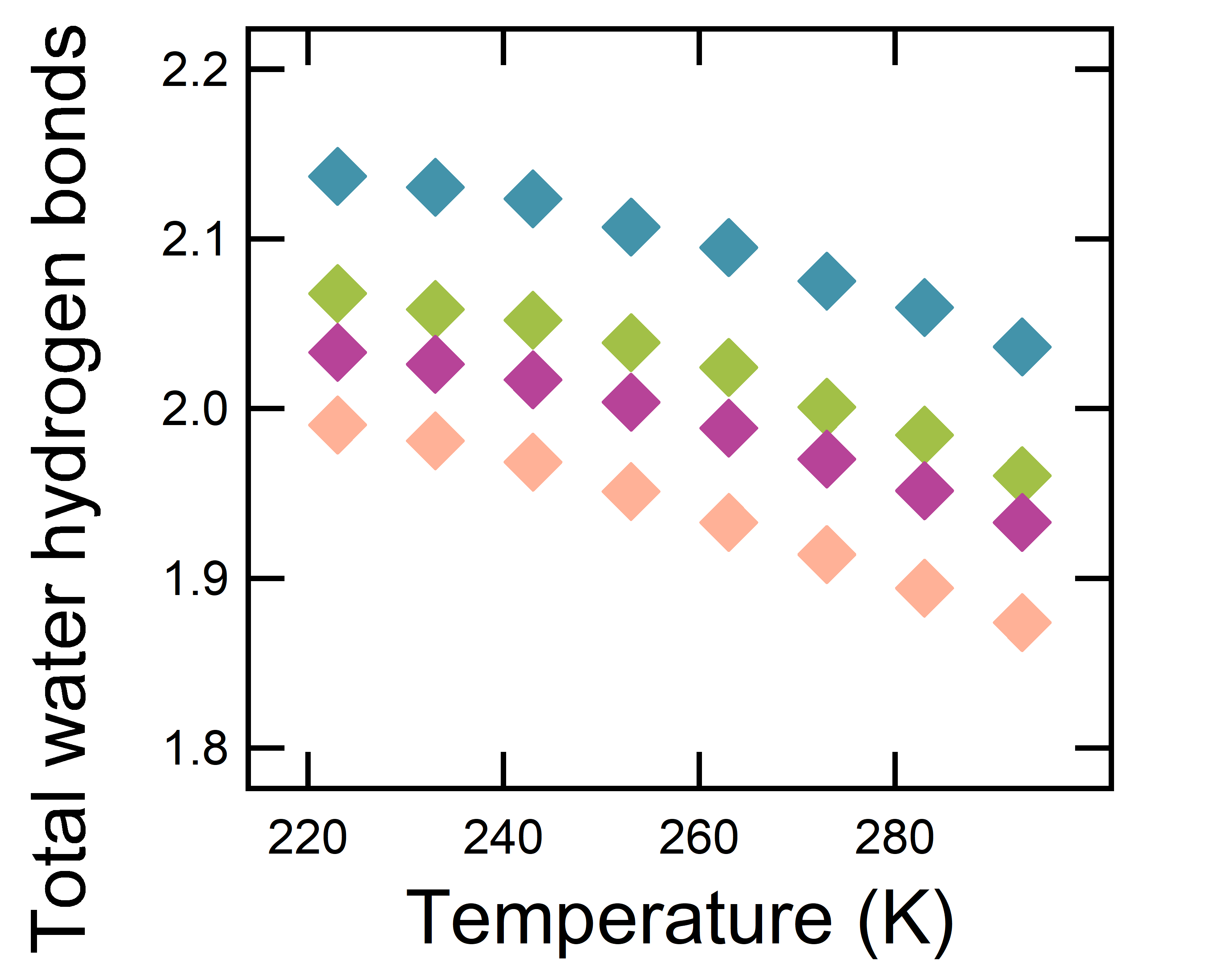}
  \caption{Temperature dependence of the average number of hydrogen bonds between water molecules or between PNIPAM and water \emph{per} water molecule, calculated over the last 100 ns. Results referring to the system with PNIPAM mass fraction of 30\%, 40\% and 60\% (w/w) are displayed in purple, green, and blue, respectively. Results obtained for bulk water are shown in pink. Errors are within symbol size.}
  \label{fgr:WWhb}
\end{figure}

We have investigated how PNIPAM affects the water structuring through the characterization of the water hydrogen bonding interactions. Figure \ref{fgr:WWhb} shows the total number of hydrogen bonds between water molecules calculated as the sum of water-water and PNIPAM-water hydrogen bonds and normalized to the number of water molecules. From the comparison with the bulk water behavior also reported in Figure \ref{fgr:WWhb} it appears that the presence of PNIPAM increases the water structuring and that the effect is more pronounced with a higher PNIPAM concentration.

\section{S6. Bulk water diffusion coefficient}
To compare with the vast literature data, we discuss here the $T$-behavior of $D_w$ for bulk water. We find that it is well described by a power-law decay, consistent with Mode Coupling Theory predictions, which gives an ideal glass transition at $T_c\sim~220 K$.  These results are in agreement with those obtained by simulations with different water models,  such as TIP4P/2005 \cite{DeMarzio2016} and SPCE \cite{camisasca2016two}. We note that such ideal glass temperature is larger than what found for other models for two reasons: (i) the TIP4P/ICE model used in this work has an overall slower diffusion coefficient than TIP4P/2005  and (ii) the path we follow is isobaric ($1atm$) as compared to isochoric paths used elsewhere. At low $T$, we find that $D_w$ is also well described by an Arrhenius fit with activation energy $\sim 80$kJ/mol, significantly larger than that observed for hydration water and again in qualitative agreement with previous simulations\cite{DeMarzio2016,camisasca2016two}.

\begin{figure}[H]
\centering
\includegraphics[width=5.5 cm]{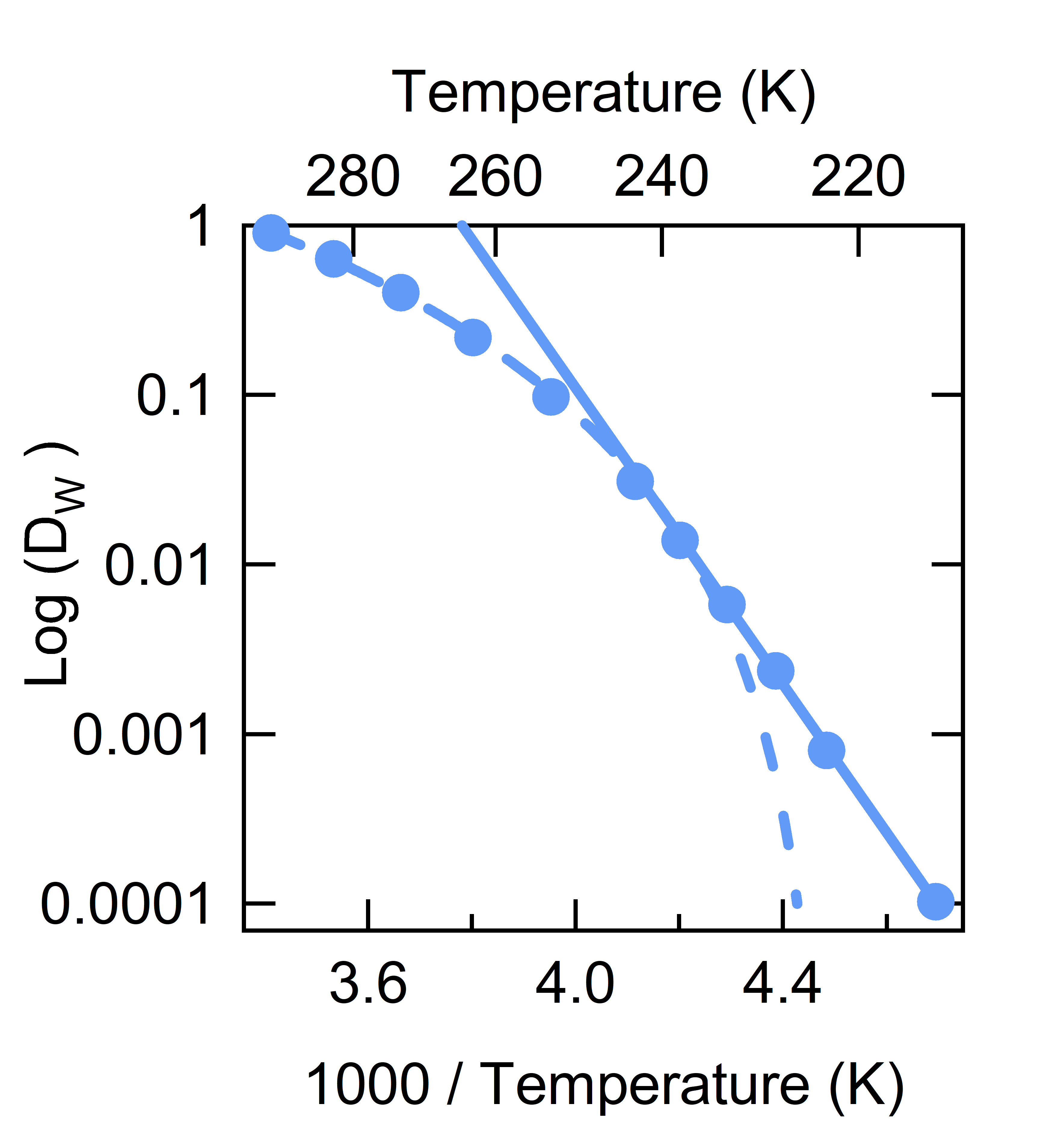}
  \caption{Temperature dependence of bulk water self-diffusion coefficient (symbols, in cm$^2/$s $\times 10^5$), which can be described by a power-law decay (dashed line) followed by an Arrhenius law (full line) at low temperatures.}
  \label{fgr:Dw}
\end{figure}

\section{S7. \emph{In silico} model of PNIPAM microgel}
A realistic all-atom model of a portion of a microgel particle was designed by cross-linking PNIPAM oligomers in the minimum conformational energy~\cite{flory1966random} through bisacrylamide junctions, as schematically shown in Figure \ref{fgr:model}. The polymeric network was built using 6 interconnected PNIPAM 30-mers each of which has a content of racemo dyads equal to 55\%. Such dyads composition corresponds to that obtained for PNIPAM synthesized without stereo-selective agents~\cite{Ray2004}, thus the model can be assumed as a representation of an atactic PNIPAM network. To mimic the percolation of the polymer scaffold in the microgel particle, extra-boundaries covalent connectivity between polymer chains was implemented. Amide groups of PNIPAM side chains were represented in \emph{trans} configuration. The resulting polymeric network (Figure \ref{fgr:model}) was hydrated by the proper shell of water molecules to set the concentration of the microgel suspension.

\begin{figure}[H]
\centering
\includegraphics[width=7 cm]{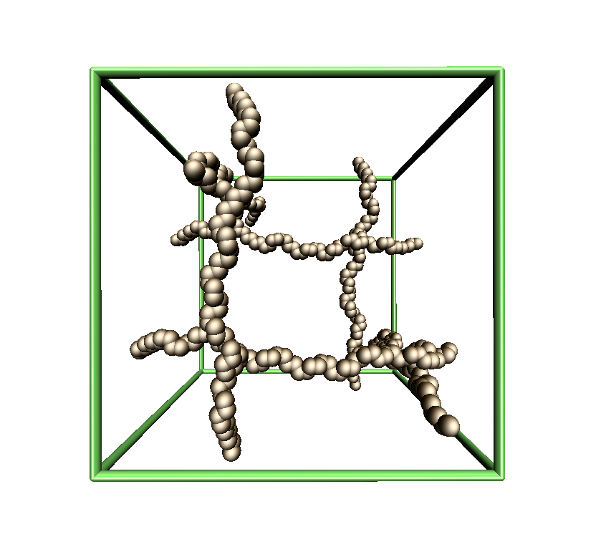}
  \caption{Schematic representation of the \emph{in silico} model of PNIPAM microgel. Only backbone heavy atoms of the polymeric network are displayed.}
  \label{fgr:model}
\end{figure}

\section{S8. Simulation protocol}
All-atom molecular dynamics simulations of microgels suspensions with PNIPAM mass fractions of 30\%, 40\% and 60\% were performed using the GROMACS 5.0.4 software~\cite{Pall2015,Abraham201519}. PNIPAM was modeled using the OPLS-AA force field~\cite{Jorgensen1996} with the implementation by Siu et al.~\cite{Siu2012}. In order to explore the supercooled temperature regime, water was described with the Tip4p/ICE model~\cite{tip4pICE}. This polymer-water force field setup was shown to reproduce the LCST transition of PNIPAM in aqueous solution~\cite{Tavagnacco2018}. MD simulations were carried out in a range of temperature between 293~K and 223~K, with a temperature step of 10 K. For each system, equilibration was first carried out at 293~K in a pressure bath at 1 bar up to a constant density value, i.e. tot-drift lower than $2 \times 10^{-3}$~g~cm$^{-3}$ over 20~ns. Pressure was controlled by the Berendsen and Parinello$-$Rahman algorithms, with time constant of 1 and 2 ps, respectively. Trajectory data were then acquired in the NVT ensemble. A similar equilibration protocol was applied at each temperature. For PNIPAM 40\% mass fraction, we also explored temperatures down to 193~K using the same procedure. In addition, to investigate the fast rotational motion of the side chain methyl groups, we extended the investigated temperature range from 183 K to 63 K with a temperature step of 20 K using an equilibration time of 70 ns and a production run of 30 ns. The leapfrog integration algorithm was employed with a time step of 2~fs. The length of bonds involving hydrogen atoms was kept fixed with the LINCS algorithm. Cubic periodic boundary conditions and minimum image convention were applied. Temperature was controlled with the velocity rescaling thermostat coupling algorithm with a time constant of 0.1~ps. Electrostatic interactions were treated with the smooth particle-mesh Ewald method with a cutoff of non-bonded interactions of 1~nm. Trajectory data were collected for 330~ns in the NVT ensemble, with a sampling of 0.2~frame/ps. The last 100~ns were typically considered for data analysis.

MD simulations were also carried out in the same temperature regime on a cubic box containing 1728 Tip4p/ICE water molecules. In this case we applied the same simulation protocol adopted for PNIPAM suspensions. Trajectory data were collected for 130~ns at each temperature.

The software MDANSE~\cite{MDANSE} was used to calculate self intermediate scattering functions. Trajectory manipulations were carried out with the software WORDOM~\cite{wordom}. The software VMD~\cite{VMD} was employed for graphical visualization.

\section{S9. Data analysis}
To evaluate the presence of global structural rearrangements as a function of the cooling down of the system, we calculated the radius of gyration of PNIPAM through the equation:
\begin{eqnarray}
R_g&=&\left(\frac{\sum_{i}\|r_i\|^2m_i}{\sum_{i}m_i}\right)^\frac{1}{2}
\label{Eq:rg}
\end{eqnarray}
where $m_i$ is the mass of the atom $i$ and $r_i$ the position of the atom $i$ with respect to the center of mass of the polymer network.

Local structural rearrangements were investigated through the analysis of the conformation and torsional dynamics of dihedral angles. We estimated the average lifetime of a rotational state $<\tau>$ which is defined as:
\begin{eqnarray}
<\tau>&=&\frac{t_{TOT} \cdot N_{DIHE}}{N_{TRANS}}
\label{Eq:dihedral}
\end{eqnarray}
where $t_{TOT}$ is the investigated time interval, $N_{TRANS}$ is the number of transitions of the dihedrals and $N_{DIHE}$ is the total number dihedral angles. A dihedral angle is classified as mobile when it performs at least one transition in the analyzed trajectory interval.
In the analysis of the torsional dynamics of methyl groups in PNIPAM side chains, we defined the dihedral of the methyl group as the angle formed by the atoms $N-C_{isopropyl}-C_{methyl}-H$. An analogous analysis was carried out for the torsional dynamics of the dihedrals formed by backbone carbon atoms.

PNIPAM$-$water and water$-$water hydrogen bonds were studied by using the geometric criteria of a donor$-$acceptor distance $(D-H \cdots A)$ lower than 0.35 nm and an angle $\theta (D-H \cdots A)$ lower than 30\degree. The dynamical behavior of the interactions was characterized by calculating the normalized intermittent time autocorrelation function which is irrespective of intervening interruptions. The corresponding hydrogen bonding lifetime was defined as the time at which the autocorrelation function is decayed of the 63\% of its amplitude.

The occurrence of hydrophobic contacts within the polymer network was accounted when the distance between methyl carbon atoms was lower than the first minimum of the corresponding radial distribution function, i.e. 0.5 nm.

The mean squared displacement of PNIPAM hydrogen atoms, $MSD$, was calculated directly from the trajectory using the following equation:
\begin{eqnarray}
MSD(t)&=&\langle |r_p(t)-r_p(0)|^2 \rangle
\label{Eq:MSD}
\end{eqnarray}
where $r_p(t)$ and $r_p(0)$ are the position vector of the PNIPAM hydrogen atom at time $t$ and $0$, with an average performed over both time origins and hydrogen atoms.

The diffusion coefficient of water molecules, $D_w$, was obtained from the long time slope of the mean squared displacement:
\begin{eqnarray}
D_w&=&\frac{1}{6}\lim_{t\to \infty} {d \over dt} \langle |r_w(t)-r_w(0)|^2 \rangle
\label{Eq:Dw}
\end{eqnarray}
where $r_w(t)$ and $r_w(0)$ are the position vector of the water oxygen atom at time $t$ and $0$, with an average performed over both time origins and water molecules. The slope was calculated over a time window of 5 ns.

The water accessible surface area, defined as van der Waals envelope of the solute molecule expanded by the radius of the solvent sphere about each solute atom centre, was evaluated for the PNIPAM microgel and lysozyme at the concentration of 67\% (w/w) using a spherical probe with a radius of 0.14 nm and the values of van der Waals radii of the work of Bondi~\cite{Bondi1964}.

When not differently specified, the errors on the calculated properties were estimated by the blocking method.

\end{document}